\documentclass[runningheads,a4paper]{llncs}

\usepackage{amssymb}
\usepackage{graphicx}

\usepackage{algorithm}
\usepackage{algorithmic}
\usepackage[colorinlistoftodos]{todonotes} 
\usepackage{amsmath}
\usepackage{multirow}
\usepackage{booktabs}

\usepackage{url}
\urldef{\mailtu}\path|patrick.styll@tuwien.ac.at|    
\newcommand{\keywords}[1]{\par\addvspace\baselineskip
\noindent\keywordname\enspace\ignorespaces#1}

\urldef{\mailsnu}\path|{dowonkim00, connectome}@snu.ac.kr|

\begin{document}

\mainmatter  

\title{Swin fMRI Transformer Predicts\\Early Neurodevelopmental Outcomes\\from Neonatal fMRI}

\titlerunning{SwiFT Predicts Neurodevelopmental Outcomes from Neonatal fMRI}

%
%

\author{Patrick Styll%
\thanks{These authors contributed equally.}%
\and Dowon Kim%
\textsuperscript{\ensuremath{\star}}
\and Jiook Cha}
\authorrunning{Patrick Styll\textsuperscript{\ensuremath{\star}}, Dowon Kim\textsuperscript{\ensuremath{\star}} and Jiook Cha}


\institute{TU Wien Informatics\\
\mailtu\\
\url{https://informatics.tuwien.ac.at/}\\
\mbox{}\\
Seoul National University\\
\mailsnu\\
\url{https://www.connectomelab.com/}}

%
%

\toctitle{Swin fMRI Transformer Predicts Early Neurodevelopmental Outcomes from Neonatal fMRI}
\tocauthor{Patrick Styll, Dowon Kim, Jiook Cha}
\maketitle

\begin{abstract}
Brain development in the first few months of human life is a critical phase characterized by rapid structural growth and functional organization. Accurately predicting developmental outcomes during this time is crucial for identifying delays and enabling timely interventions. This study introduces the SwiFT (Swin 4D fMRI Transformer) model, designed to predict Bayley-III composite scores using neonatal fMRI from the Developing Human Connectome Project (dHCP). To enhance predictive accuracy, we apply dimensionality reduction via group independent component analysis (ICA) and pretrain SwiFT on large adult fMRI datasets to address the challenges of limited neonatal data. Our analysis shows that SwiFT significantly outperforms baseline models in predicting cognitive, motor, and language outcomes, leveraging both single-label and multi-label prediction strategies. The model’s attention-based architecture processes spatiotemporal data end-to-end, delivering superior predictive performance. Additionally, we use Integrated Gradients with Smoothgrad sQuare (IG-SQ) to interpret predictions, identifying neural spatial representations linked to early cognitive and behavioral development. These findings underscore the potential of Transformer models to advance neurodevelopmental research and clinical practice.
\end{abstract}

\section{Introduction}

Brain development during the first few months of life is a period of rapid structural and functional reorganization, making it a critical window for identifying potential neurodevelopmental deficits. Accurately predicting developmental outcomes during this period is essential for enabling early interventions that can mitigate the lifelong impact of developmental delays. Neonatal fMRI data, such as those from the Developing Human Connectome Project (dHCP), have shown potential for predicting neurodevelopmental outcomes~\cite{LI2024114168}. However, the spatiotemporal complexity of neonatal brain activity presents significant challenges for conventional analysis methods.
This study investigates the potential of the Swin 4D fMRI Transformer (SwiFT)~\cite{kim2023swiftswin4dfmri}, a deep learning architecture designed to process high-dimensional fMRI data, for predicting neurodevelopmental outcomes from neonatal fMRI. Unlike existing methods, SwiFT leverages 4D spatiotemporal attention mechanisms to effectively capture dynamic brain connectivity patterns, offering a novel approach for analyzing neonatal fMRI data. Specifically, we aim to predict Bayley Scales of Infant and Toddler Development, Third Edition (Bayley-III/BSID-III) composite scores, encompassing cognitive, language, and motor skills, using neonatal fMRI from the dHCP dataset. To address the challenges of limited neonatal data and high dimensionality, we explore dimensionality reduction using group Independent Component Analysis (ICA) and pretraining SwiFT on large publicly available adult fMRI datasets. 
We hypothesize that (1) SwiFT outperforms baseline models in predicting Bayley-III scores due to its ability to capture complex spatiotemporal patterns in fMRI data and (2) incorporating ICA features and pretraining further improves predictive accuracy. Furthermore, we aim to interpret SwiFT's predictions using Integrated Gradients with Smoothgrad sQuare (IG-SQ)~\cite{captum1} to identify brain regions associated with early cognitive, language, and motor development. 
This research advances the state of neurodevelopmental prediction by introducing a robust and interpretable framework for neonatal fMRI analysis. By addressing key methodological challenges, our findings lay the groundwork for future applications in neurodevelopmental research and clinical practice.

In Section~\ref{sec:methods}, we present the dataset, its acquisition, processing, target values, and an overview of SwiFT. Section~\ref{sec:experiments} outlines our experimental settings, including (i) baselines, (ii) SwiFT on raw fMRI data for single- and multi-label prediction, and (iii) SwiFT on group-ICA features for the same tasks. Section~\ref{sec:interpretations} focuses on IG-SQ analysis of the best model, interpreting IG maps for delays in cognitive, language, and motor skills. Finally, Section~\ref{sec:conclusion} discusses limitations and future directions.


\section{Methods}
\label{sec:methods}

\paragraph{Target Developmental Outcomes}
This study focuses on Bayley Scales of Infant and Toddler Development, Third Edition (Bayley-III), composite scores in cognitive, language, and motor domains. Developmental delays are categorized based on established thresholds ~\cite{johnson2014using}:

\begin{itemize}
    \item greater than or equal to 85: average development
    \item lower than 85: risk of developmental delay
    \item lower than 75: moderate to severe mental impairment
\end{itemize}

Following prior studies~\cite{he2020multi}, we simplify the classification task into a binary format, with a threshold of 85 distinguishing low- and high-risk cases. Regression tasks are also conducted to predict continuous Bayley-III scores.

\paragraph{Dataset}
The Developing Human Connectome Project (dHCP) dataset provides a rich basis for developing predictive models of early neurodevelopment. It includes 783 neonates born between 23–44 gestational weeks, with imaging performed between 26–45 post-menstrual weeks. Resting-state fMRI (rs-fMRI) data were acquired using a 3T Philips Achieva scanner and preprocessed via the dHCP pipeline for motion correction, denoising, and spatial normalization~\cite{FITZGIBBON2020117303}.
Bayley-III scores in cognition, language, and motor domains that were collected in follow-up assessment, planned for 18 months corrected age but affected by the COVID-19 pandemic (median: $18 \text{ months} + 12 \text{ days}$, range $17 + 8$ -- $34 + 15$), were available for 739 infants, with 619 of them also having usable fMRI data. The composite scores showed a mean close to 100 with imbalanced risks of developmental delay: $\approx5\%$ for cognition and motor domains and $\approx18\%$ for language. Strong positive correlations were observed between cognition, language, and motor scores (63.08\%, 51.94\%, and 57.27\%, respectively). Maternal factors (e.g., smoking) showed no significant correlation with scores.

\begin{figure}[h]
    \centering
    \includegraphics[width=.7\columnwidth]{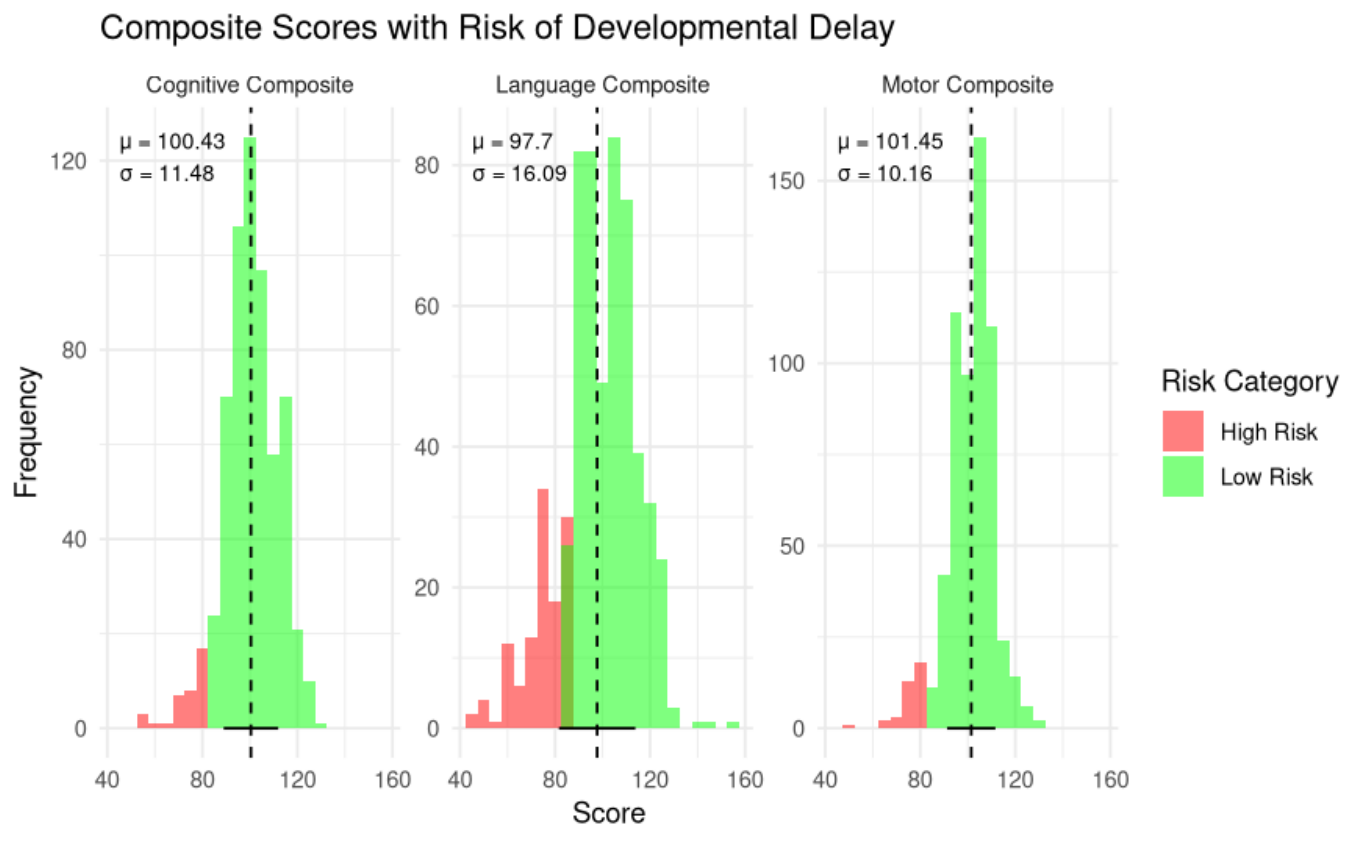}
    \caption{Distributions of composite scores based on binary classification. Assumptions made in~\cite{johnson2014using} largely hold true.}
    \label{fig:bsid-demographics}
\end{figure}


\paragraph{MRI Acquisition and Processing}  
MRI data were acquired on a 3T Philips Achieva system with a neonatal 32-channel head coil. The imaging protocol was tailored to the unique characteristics of the neonatal brain. Resting-state fMRI (rs-fMRI) was captured with a multiband-accelerated echo-planar imaging sequence (TR: 392 ms, TE: 38 ms, resolution: $2.15\text{mm}^3$) over 15 minutes (2300 volumes). Processing followed the dHCP pipeline, which is designed to address the challenges of neonatal imaging. This included motion and distortion correction, skull stripping, ICA-based denoising with FIX, and high-pass temporal filtering to remove slow drifts.
To align the data to a neonatal-specific template space, we adopted a pipeline inspired by the automated resting-state functional framework developed for the dHCP dataset~\cite{FITZGIBBON2020117303}. Specifically, we utilized the 40-week T1-weighted template from the extended dHCP Augmented Volumetric Atlas~\cite{Schuh2018}, downsampled from a spatial resolution of 0.5~$mm^3$ isotropic to 2.15~$mm^3$ isotropic to match the native resolution of the fMRI images. Image registration was performed using ANTs~\cite{tustison_antsx_2021} on the mean image of all timepoints, with the resulting transformation applied independently to each timeframe in parallel. The processed outputs served as inputs for the subsequent experiments.

\paragraph{Swin fMRI Transformer}  
To predict developmental outcomes from neonatal fMRI, we used SwiFT, a Transformer-based model designed for high-dimensional spatiotemporal fMRI data. SwiFT processes 4D fMRI volumes by segmenting them into patches, embedding them as tokens, and passing them through layers of 4D Swin Transformer blocks. These blocks utilize a shifted window multi-head self-attention mechanism to capture both local and global spatiotemporal dependencies while minimizing computational costs through patch merging. SwiFT demonstrated robust performance on diverse tasks (e.g., predicting sex, age, intelligence, or psychiatric diagnosis), showcasing its effectiveness in learning spatiotemporal patterns from the human brains~\cite{kim2023swiftswin4dfmri}.

\section{Experiments}
\label{sec:experiments}

\subsection{Experimental Settings}
\label{sec:experimental_settings}

All experiments were carried out on the \emph{Perlmutter} supercomputer, an HPE Cray EX system. A holdout strategy with a 70:15:15 data split was used for training, validation, and testing. To address data imbalance (see Section~\ref{sec:methods}), stratified splits were applied based on the BSID-III scores. For benchmarking, 5-fold cross-validation ensured consistent and fair splits between experiments.
For binary classification, we used weighted focal loss~\cite{lin2018focallossdenseobject} to prioritize harder-to-classify (\emph{at-risk}) cases. This approach significantly improved performance, especially for these challenging instances.
Following ~\cite{kim2023swiftswin4dfmri}, we padded input images to 64x64x64 to align with the model architecture, adapting from the 96x96x96 padding used in ~\cite{kim2023swiftswin4dfmri} due to our native image size of 45x55x45.
For classification tasks, we evaluated performance using the area under the ROC curve (AUC), accuracy, and balanced accuracy (ACC$_{bal}$) to account for class imbalance. For regression tasks, mean absolute error (MAE) and mean squared error (MSE) were used, along with adjusted metrics (MAE$_{adj}$ and MSE$_{adj}$) to reflect re-scaled values for interpretability.

\subsection{Baselines}

We established baseline performance using ROI-based deep learning models, including BrainNetCNN \cite{KAWAHARA20171038}, VanillaTF, and Brain Network Transformer (BNT) \cite{KAN2022}, which take functional connectivity matrices as input (computed via Pearson correlations of 87 brain regions in the dHCP neonatal brain atlas). Hyper-parameters and implementations followed \cite{KAN2022}. Additionally, we included XGBoost \cite{CHEN2016} as a traditional machine learning baseline, using the flattened upper triangular correlation matrix as input. Experiments for both classification and regression can be seen in Table~\ref{tab:baseline-models}.


\begin{table}[h]
\centering
\caption{Baseline performances on regression (left) and classification (right).}
\label{tab:baseline-models}

\begin{minipage}{0.45\textwidth}
\resizebox{\textwidth}{!}{%
\begin{tabular}{|l|c|c|c|}
\hline
\multirow{2}{*}{Model} & \multicolumn{3}{c|}{Cog} \\ \cline{2-4} 
 & $\text{MSE}_{\text{adj}}$ & MAE & $\text{MAE}_{\text{adj}}$ \\ \hline
XGBoost & $155.6_{\pm15.0}$ & $\mathbf{0.84_{\pm0.03}}$ & $\mathbf{9.59_{\pm0.48}}$ \\ \hline
BrainNetCNN & $\mathbf{113.6_{\pm54.2}}$ & $0.84_{\pm0.09}$ & $9.66_{\pm1.55}$ \\ \hline
VanillaTF & $210.1_{\pm91.4}$ & $0.91_{\pm0.04}$ & $10.43_{\pm1.13}$ \\ \hline
BNT & $147.1_{\pm86.7}$ & $0.87_{\pm0.07}$ & $10.01_{\pm1.33}$ \\ \hline
\end{tabular}
}

\vspace{.3em}

\resizebox{\textwidth}{!}{%
\begin{tabular}{|l|c|c|c|}
\hline
\multirow{2}{*}{Model} & \multicolumn{3}{c|}{Lang} \\ \cline{2-4} 
 & $\text{MSE}_{\text{adj}}$ & MAE & $\text{MAE}_{\text{adj}}$ \\ \hline
XGBoost & $304.7_{\pm33.9}$ & $\mathbf{0.86_{\pm0.07}}$ & $\mathbf{13.94_{\pm0.75}}$ \\ \hline
BrainNetCNN & $317.6_{\pm109.6}$ & $0.88_{\pm0.04}$ & $14.28_{\pm1.31}$ \\ \hline
VanillaTF & $\mathbf{279.7_{\pm96.3}}$ & $0.91_{\pm0.03}$ & $14.77_{\pm1.21}$ \\ \hline
BNT & $355.1_{\pm49.6}$ & $0.90_{\pm0.03}$ & $14.68_{\pm1.20}$ \\ \hline
\end{tabular}
}

\vspace{.3em}

\resizebox{\textwidth}{!}{%
\begin{tabular}{|l|c|c|c|}
\hline
\multirow{2}{*}{Model} & \multicolumn{3}{c|}{Mot} \\ \cline{2-4} 
 & $\text{MSE}_{\text{adj}}$ & MAE & $\text{MAE}_{\text{adj}}$ \\ \hline
XGBoost & $124.9_{\pm18.0}$ & $\mathbf{0.81_{\pm0.05}}$ & $\mathbf{8.55_{\pm0.59}}$ \\ \hline
BrainNetCNN & $\mathbf{115.74_{\pm28.5}}$ & $0.84_{\pm0.04}$ & $8.93_{\pm0.75}$ \\ \hline
VanillaTF & $173.0_{\pm108.2}$ & $0.86_{\pm0.03}$ & $9.11_{\pm0.84}$ \\ \hline
BNT & $153.7_{\pm100.4}$ & $0.81_{\pm0.04}$ & $8.58_{\pm0.90}$ \\ \hline
\end{tabular}
}
\end{minipage}
\hspace{3em}
\begin{minipage}{0.4\textwidth}

\resizebox{\textwidth}{!}{%
\begin{tabular}{|l|c|c|c|}
\hline
\multirow{2}{*}{Model} & \multicolumn{3}{c|}{Cog} \\ \cline{2-4} 
 & $\text{ACC}_{\text{bal}}$ & ACC & AUC \\ \hline
XGBoost & $50.0_{\pm0.0}$ & $\mathbf{93.8_{\pm0.4}}$ & $51.5_{\pm10.4}$ \\ \hline
BrainNetCNN & $49.6_{\pm0.4}$ & $93.0_{\pm1.0}$ & $47.5_{\pm4.8}$ \\ \hline
VanillaTF & $\mathbf{52.2_{\pm4.2}}$ & $92.5_{\pm1.0}$ & $55.5_{\pm9.5}$ \\ \hline
BNT & $51.2_{\pm3.7}$ & $93.1_{\pm1.0}$ & $\mathbf{59.8_{\pm4.2}}$ \\ \hline
\end{tabular}
}

\vspace{.3em}

\resizebox{\textwidth}{!}{%
\begin{tabular}{|l|c|c|c|}
\hline
\multirow{2}{*}{Model} & \multicolumn{3}{c|}{Lang} \\ \cline{2-4} 
 & $\text{ACC}_{\text{bal}}$ & ACC & AUC \\ \hline
XGBoost & $50.0_{\pm0.9}$ & $\mathbf{79.5_{\pm0.8}}$ & $50.4_{\pm7.5}$ \\ \hline
BrainNetCNN & $50.3_{\pm3.0}$ & $77.5_{\pm3.1}$ & $\mathbf{56.1_{\pm4.9}}$ \\ \hline
VanillaTF & $49.9_{\pm2.4}$ & $74.0_{\pm2.4}$ & $54.1_{\pm6.8}$ \\ \hline
BNT & $\mathbf{51.1_{\pm3.9}}$ & $74.8_{\pm3.1}$ & $\mathbf{57.1_{\pm5.6}}$ \\ \hline
\end{tabular}
}

\vspace{.3em}

\resizebox{\textwidth}{!}{%
\begin{tabular}{|l|c|c|c|}
\hline
\multirow{2}{*}{Model} & \multicolumn{3}{c|}{Mot} \\ \cline{2-4} 
 & $\text{ACC}_{\text{bal}}$ & ACC & AUC \\ \hline
XGBoost & $\mathbf{49.7_{\pm0.3}}$ & $\mathbf{93.1_{\pm0.5}}$ & $51.2_{\pm6.2}$ \\ \hline
BrainNetCNN & $49.5_{\pm0.7}$ & $92.7_{\pm1.3}$ & $\mathbf{57.1_{\pm5.6}}$ \\ \hline
VanillaTF & $49.2_{\pm0.4}$ & $92.1_{\pm0.8}$ & $47.4_{\pm7.9}$ \\ \hline
BNT & $49.5_{\pm0.3}$ & $92.7_{\pm0.7}$ & $44.7_{\pm6.8}$ \\ \hline
\end{tabular}
}
\end{minipage}

\end{table}


\subsection{fMRI Volumes}
\label{sec:raw_fmri}

We evaluated the SwiFT model's performance on fMRI data in an end-to-end manner, focusing on single-label and multi-label prediction tasks. Multi-label prediction was hypothesized to enhance performance due to strong correlations among Bayley scores (see Section~\ref{sec:methods}), as supported by prior work~\cite{He2020}.
The models were fine-tuned across a predefined hyperparameter search space, including data augmentation strategies (affine-only, intensity-only, and combined transformations), learning rates, batch sizes, and patch sizes. Sequence length of the input was varied to evaluate its impact: a baseline length of 20 was used per ~\cite{kim2023swiftswin4dfmri}, with additional experiments using lengths of 50 and 100 to assess temporal dynamics, as longer sequences have shown promise in prior studies for predicting intelligence in young adults and elders~\cite{kim2023swiftswin4dfmri}.
To leverage publicly available fMRI datasets, we pre-trained SwiFT using contrastive self-supervised learning~\cite{dave2022TCLR} on data from the UK Biobank (UKB)~\cite{alfaro2018UKBiobank}~\cite{miller2016UKBiobank}, Human Connectome Project (HCP)~\cite{vanEssen2013HCP}, and Adolescent Brain Cognitive Development (ABCD) study~\cite{casey2018ABCD}. While this approach utilizes larger datasets, differences between neonatal and adult or adolescent brain data present potential limitations. To address size discrepancies, neonatal images were padded to 96x96x96, contrasting with experiments trained from scratch, where images were padded to 64x64x64 (see Section~\ref{sec:experimental_settings}).


\begin{figure}[h]
    \centering
    \includegraphics[width=1\columnwidth]{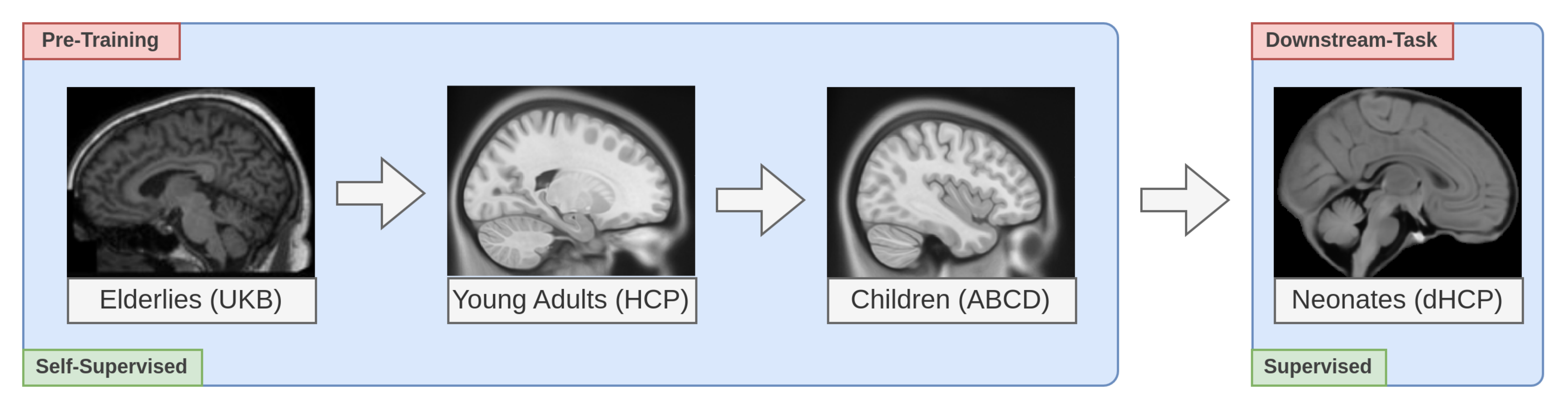}
    \caption{Rationale of using large amounts of brain fMRI data across different datasets to make up for the little amount of data available for neonates. We expect that learned features from elderly, adult and child data will generalize to limited neonatal fMRI data and thus improve downstream performance for neonates.}
    \label{fig:pretraining}
\end{figure}


\paragraph{Single-Label Prediction}

The choice of sequence length had a notable impact on the model's performance. A length of 50 yielded the best results, while shorter (20) or longer (100) sequences performed slightly worse, suggesting that an optimal balance is necessary.
As shown in Table~\ref{tab:bsid-raw}, pre-training resulted in marginal improvements in predictive power for BSID-III scores. However, increased padding might have negatively influenced the performance. Compared to baseline models, SwiFT yielded significant performance gains for both regression and classification tasks. For cognitive and language predictions, adjusted MAE improved by approximately 11\%, while gains for motor regression were minimal. In classification, balanced accuracy ($\text{ACC}_{bal}$) improved by 5-7\% points, exceeding guessing levels.

\begin{table}[h]
\centering
\caption{SwiFT’s performance with raw fMRI as input on single-label (left) and multi-label (right) predictions.}
\label{tab:bsid-raw}

\begin{minipage}{0.4\textwidth}
\resizebox{\textwidth}{!}{%
\begin{tabular}{|l|c|c|c|}
\hline
\multirow{2}{*}{Method} & \multicolumn{3}{c|}{Cog} \\ \cline{2-4} 
 & $\text{MSE}_{adj}$ & MAE & $\text{MAE}_{adj}$ \\ \hline
Scratch & $127.8_{\pm11.8}$ & $0.77_{\pm0.03}$ & $8.8_{\pm0.7}$ \\\hline
Pre-Tr. & $\mathbf{123.5_{\pm12.6}}$ & $\mathbf{0.76_{\pm0.02}}$ & $\mathbf{8.6_{\pm0.5}}$ \\ \hline
\hline
 & ACC & $\text{ACC}_{bal}$ & AUC \\ \hline
Scratch & $94.4_{\pm0.7}$ & $\mathbf{54.9_{\pm3.8}}$ & $\mathbf{56.9_{\pm5.2}}$ \\ \hline
Pre-Tr. & $\mathbf{94.9_{\pm1.2}}$ & $54.1_{\pm3.6}$ & $54.6_{\pm3.2}$ \\
\hline
\end{tabular}
}

\vspace{.3em}

\resizebox{\textwidth}{!}{%
\begin{tabular}{|l|c|c|c|}
\hline
\multirow{2}{*}{Method} & \multicolumn{3}{c|}{Lang} \\ \cline{2-4} 
 & $\text{MSE}_{adj}$ & MAE & $\text{MAE}_{adj}$ \\ \hline
Scratch & $293.4_{\pm20.1}$ & $0.79_{\pm0.04}$ & $12.9_{\pm0.5}$ \\ \hline
Pre-Tr. & $\mathbf{287.4_{\pm15.4}}$ & $\mathbf{0.78_{\pm0.02}}$ & $\mathbf{12.6_{\pm0.3}}$ \\ \hline
\hline
 & ACC & $\text{ACC}_{bal}$ & AUC \\ \hline
Scratch & $82.7_{\pm2.8}$ & $56.2_{\pm2.1}$ & $\mathbf{56.6_{\pm3.2}}$ \\ \hline
Pre-Tr. & $\mathbf{83.9_{\pm1.4}}$ & $\mathbf{57.6_{\pm3.1}}$ & $56.3_{\pm3.1}$ \\ \hline
\end{tabular}
}

\vspace{.3em}

\resizebox{\textwidth}{!}{%
\begin{tabular}{|l|c|c|c|}
\hline
\multirow{2}{*}{Method} & \multicolumn{3}{c|}{Mot} \\ \cline{2-4} 
 & $\text{MSE}_{adj}$ & MAE & $\text{MAE}_{adj}$ \\ \hline
Scratch & $\mathbf{103.1_{\pm9.2}}$  & $\mathbf{0.81_{\pm0.08}}$  & $\mathbf{8.4_{\pm0.9}}$ \\ \hline
Pre-Tr. & $105.7_{\pm9.8}$  & $0.82_{\pm0.09}$  & $8.6_{\pm1.1}$ \\ \hline
\hline
 & ACC & $\text{ACC}_{bal}$ & AUC \\ \hline
Scratch & $\mathbf{93.6_{\pm0.6}}$  & $54.2_{\pm3.2}$  & $55.5_{\pm3.7}$ \\ \hline
Pre-Tr. & $92.8_{\pm1.2}$  & $\mathbf{55.1_{\pm3.3}}$  & $\mathbf{55.7_{\pm3.9}}$ \\ \hline
\end{tabular}
}
\end{minipage}
\hspace{3em}
\begin{minipage}{0.4\textwidth}
\resizebox{\textwidth}{!}{%
\begin{tabular}{|l|c|c|c|}
\hline
\multirow{2}{*}{Method} & \multicolumn{3}{c|}{Cog} \\ \cline{2-4} 
 & $\text{MSE}_{adj}$ & MAE & $\text{MAE}_{adj}$ \\ \hline
Scratch & $125.7_{\pm16.6}$ & $\mathbf{0.64_{\pm0.07}}$ & $\mathbf{8.5_{\pm0.6}}$ \\\hline
Pre-Tr. & $\mathbf{119.2_{\pm12.6}}$ & $0.64_{\pm0.12}$ & $8.6_{\pm1.2}$ \\ \hline
\hline
 & ACC & $\text{ACC}_{bal}$ & AUC\\ \hline
Scratch & $\mathbf{96.4_{\pm1.0}}$ & $59.7_{\pm2.1}$ & $62.6_{\pm4.0}$ \\ \hline
Pre-Tr. & $95.6_{\pm2.4}$ & $\mathbf{66.0_{\pm2.9}}$ & $\mathbf{62.7_{\pm4.7}}$ \\
\hline
\end{tabular}
}

\vspace{.3em}

\resizebox{\textwidth}{!}{%
\begin{tabular}{|l|c|c|c|}
\hline
\multirow{2}{*}{Method} & \multicolumn{3}{c|}{Lang} \\ \cline{2-4} 
 & $\text{MSE}_{adj}$ & MAE & $\text{MAE}_{adj}$ \\ \hline
Scratch & $212.4_{\pm17.7}$ & $0.87_{\pm0.05}$ & $11.6_{\pm0.7}$ \\ \hline
Pre-Tr. & $\mathbf{205.3_{\pm15.3}}$ & $\mathbf{0.86_{\pm0.07}}$ & $\mathbf{11.5_{\pm0.9}}$ \\ \hline
\hline
 & ACC & $\text{ACC}_{bal}$ & AUC \\ \hline
Scratch & $83.3_{\pm1.5}$ & $61.7_{\pm1.7}$ & $62.0_{\pm1.9}$ \\ \hline
Pre-Tr. & $\mathbf{83.7_{\pm1.3}}$ & $\mathbf{62.3_{\pm1.9}}$ & $\mathbf{63.2_{\pm2.2}}$ \\ \hline
\end{tabular}
}

\vspace{.3em}

\resizebox{\textwidth}{!}{%
\begin{tabular}{|l|c|c|c|}
\hline
\multirow{2}{*}{Method} & \multicolumn{3}{c|}{Mot} \\ \cline{2-4} 
 & $\text{MSE}_{adj}$ & MAE & $\text{MAE}_{adj}$ \\ \hline
Scratch & $\mathbf{99.8_{\pm15.3}}$  & $\mathbf{0.56_{\pm0.03}}$  & $\mathbf{7.4_{\pm0.5}}$ \\ \hline
Pre-Tr. & $103.0_{\pm13.2}$  & $0.57_{\pm0.04}$  & $7.6_{\pm0.6}$ \\ \hline
\hline
 & ACC & $\text{ACC}_{bal}$ & AUC \\ \hline
Scratch & $95.2_{\pm0.7}$  & $\mathbf{57.8_{\pm2.1}}$  & $\mathbf{59.7_{\pm1.7}}$ \\ \hline
Pre-Tr. & $\mathbf{95.2_{\pm0.5}}$ & $56.3_{\pm1.9}$  & $58.9_{\pm1.9}$ \\ \hline
\end{tabular}
}
\end{minipage}

\end{table}

\paragraph{Multi-Label Prediction}

To predict all three classes simultaneously, we used a shared classification/regression head, a design choice based on the expected correlation among these values (see Section~\ref{sec:methods}).
Changes in sequence length showed similar effects to single-label predictions. As shown in Table~\ref{tab:bsid-raw}, pre-training provided marginal improvements, highlighting the limited transferability of adult data to neonatal predictions. However, multi-label regression results align with previous studies~\cite{He2020}, showing performance gains across all classes for both regression and classification. These improvements may stem from shared learning across developmental domains, enabling the model to capture complex, interrelated features of early brain development.

\subsection{Dimensionality Reduction via Group ICA}

Group Independent Component Analysis (Group ICA)~\cite{Hyvarinen1999}~\cite{Beckmann2004} is a widely used data-driven technique in neuroimaging that decomposes fMRI data into independent spatial and temporal components. These components represent distinct neural processes or functional networks. Group ICA reduces the dimensionality of high-dimensional fMRI data while preserving significant information and isolating biologically interpretable brain networks.
In this study, we use Group ICA features as inputs. These features encapsulate complex brain activity patterns in a compact representation, enhancing expressiveness and facilitating brain network-level analysis. We hypothesize that this approach will improve model interpretability and predictive performance by leveraging the reduced dimensionality and biologically relevant features.

\subsubsection{Preparation}
\paragraph{Data Sampling}
\label{sec:data-sampling}

To generate the Group ICA map, we selected a subset of normally developing neonates, defined as those with cognitive, language, and motor BSID-III scores above 85, indicating low risk of developmental delay. Only normally developing data were used to reduce variability and noise, ensuring a more reliable feature set.
Out of 725 assessed subjects, 431 met the criteria for normal development, with 358 having usable fMRI data. To further enhance stability, we excluded neonates with gestational ages outside 36 to 44 weeks, as defined by the dHCP volumetric atlas~\cite{Schuh2018}, resulting in a final cohort of 348 subjects. From this group, we randomly selected 100 neonates for Group ICA analysis.

\paragraph{Group ICA and Dimensionality Reduction}

\begin{figure}[h]
    \centering
    \includegraphics[width=0.5\columnwidth]{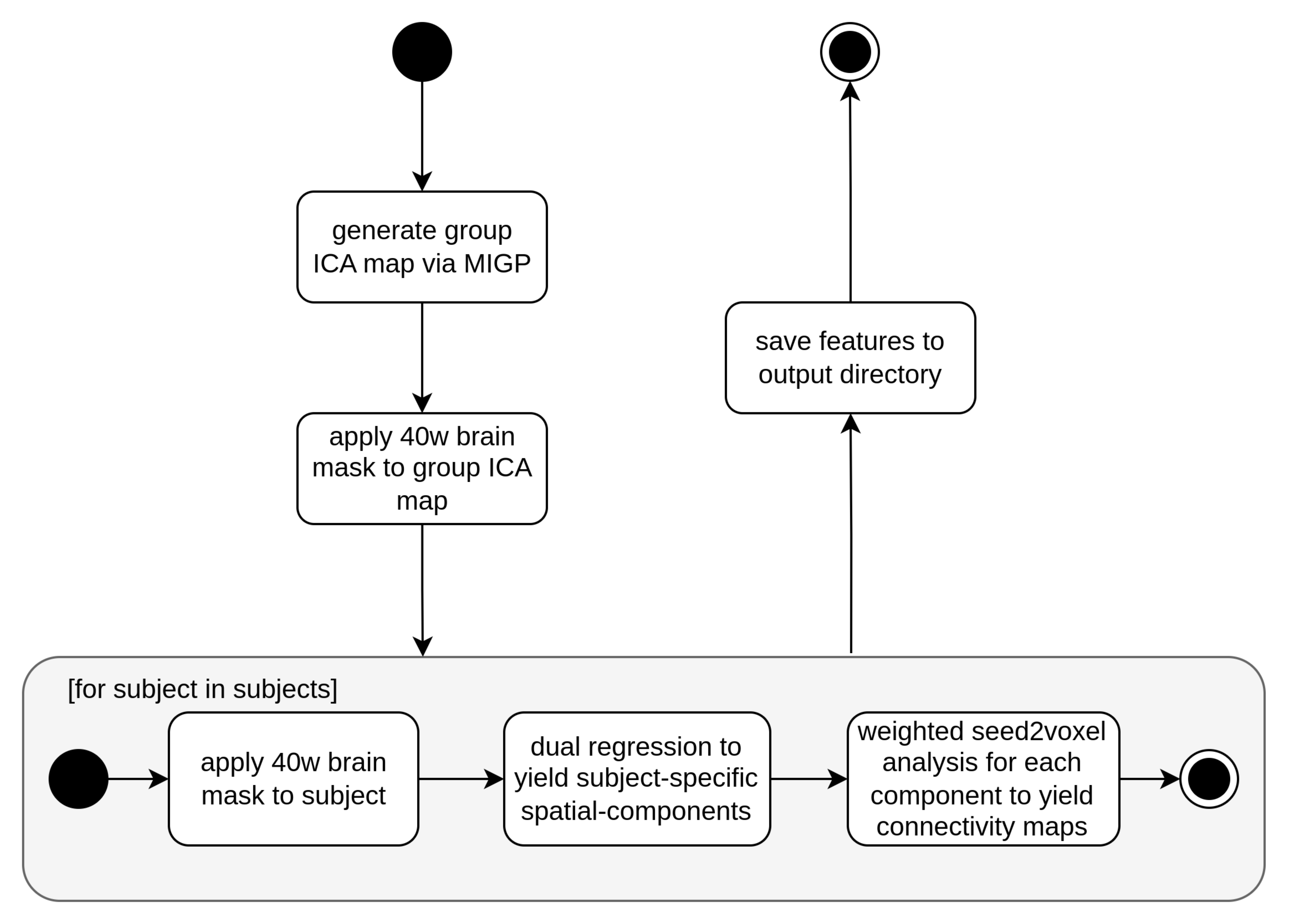}
    \caption{UML Flowchart depicting the workflow to extract features and connectivity maps by leveraging Group Independent Component Analysis ~\cite{ALFAROALMAGRO2018400},~\cite{Miller2016} and~\cite{GAL2022118920}.}
    \label{fig:ica-workflow}
\end{figure}

The workflow for Group ICA and functional connectivity analysis (see Figure~\ref{fig:ica-workflow}) follows methodologies established for the UKB dataset~\cite{ALFAROALMAGRO2018400}~\cite{Miller2016}. For the selected 100 subjects, we used MIGP (MELODIC's Incremental Group-PCA) to extract the top 1200 weighted spatial eigenvectors, approximating traditional PCA while remaining computationally efficient for large datasets~\cite{Smith2014}. Spatial Independent Component Analysis (spatial-ICA) was then applied using FSL MELODIC~\cite{Hyvarinen1999}~\cite{Beckmann2004} to parcellate the data into spatially independent components (ICs). We evaluated dimensionalities of 25 and 100 ICs, consistent with prior work on the UKB dataset~\cite{ALFAROALMAGRO2018400}~\cite{Miller2016}~\cite{SMITH2013144}~\cite{doi}. Higher dimensionalities result in smaller, more numerous regions within spatial maps. Since FIX denoising was performed during preprocessing, we did not identify artifactual components.
Using the generated group ICA maps, we performed dual regression to obtain subject-specific spatial maps for each IC. Dual regression comprises two steps: (i) spatial regression, projecting group-level IC maps onto individual data, and (ii) temporal regression, deriving subject-specific timeseries for each IC. To further explore functional connectivity, we conducted seed-to-voxel analysis, correlating timeseries of brain regions with those of all other voxels. The resulting whole-brain functional connectivity maps served as input features for the SwiFT model to predict neurodevelopmental outcomes.

\subsubsection{Results}

\paragraph{Training Process}

For each dimensionality, the model was tuned using a predefined hyperparameter search space, including learning rate, patch size, and data augmentation strategies (see Section~\ref{sec:raw_fmri}). With ICA reducing the input data to a fraction of its original dimensionality, careful tuning was critical to ensure effective convergence. Given the data demands of transformer architectures and the reduced training data size, we increased the maximum training epochs to 100. To mitigate overfitting, we applied data augmentation techniques such as affine and intensity transformations to enhance sample diversity and improve generalizability. Due to the specific sequence lengths required for IC-based training (42 and 100), transfer learning was not utilized in these experiments.
We addressed potential data leakage by ensuring that the 100 healthy subjects used for Group ICA were always included in the training set, avoiding overlap with validation or test sets. This step minimized bias and maintained the reliability of our results. Experimental outcomes are summarized in Tables \ref{tab:bsid-ica}, presenting the average test performance and standard deviations of the fine-tuned models.

\begin{table}[h]
\centering
\caption{SwiFT’s performance with IC-extracted features as input on single-label (left) and multi-label (right) predictions.}
\label{tab:bsid-ica}

\begin{minipage}{0.4\textwidth}
\resizebox{\textwidth}{!}{%
\begin{tabular}{|l|c|c|c|}
\hline
\multirow{2}{*}{ICs} & \multicolumn{3}{c|}{Cog} \\ \cline{2-4} 
 & $\text{MSE}_{adj}$ & MAE & $\text{MAE}_{adj}$ \\ \hline
25 & $127.2_{\pm6.9}$  & $\mathbf{0.74_{\pm0.03}}$  & $\mathbf{8.7_{\pm0.4}}$ \\ \hline
100 & $\mathbf{126.8_{\pm8.2}}$  & $0.74_{\pm0.07}$ & $8.7_{\pm0.9}$ \\ \hline
\hline
 & ACC & $\text{ACC}_{bal}$ & AUC \\ \hline
25 & $93.7_{\pm1.5}$ & $\mathbf{55.7_{\pm4.0}}$ & $\mathbf{56.6_{\pm4.7}}$ \\ \hline
100 & $\mathbf{93.8_{\pm1.3}}$ & $52.4_{\pm1.2}$  & $53.1_{\pm2.4}$ \\ \hline
\end{tabular}
}

\vspace{.3em}

\resizebox{\textwidth}{!}{%
\begin{tabular}{|l|c|c|c|}
\hline
\multirow{2}{*}{ICs} & \multicolumn{3}{c|}{Lang} \\ \cline{2-4} 
 & $\text{MSE}_{adj}$ & MAE & $\text{MAE}_{adj}$ \\ \hline
25 & $\mathbf{293.4_{\pm11.4}}$ & $0.82_{\pm0.02}$  & $12.6_{\pm0.3}$ \\ \hline
100 & $299.5_{\pm13.3}$ & $\mathbf{0.8_{\pm0.02}}$ & $\mathbf{12.4_{\pm0.3}}$ \\ \hline
\hline
 & ACC & $\text{ACC}_{bal}$ & AUC \\ \hline
25 & $\mathbf{84.0_{\pm2.9}}$ & $\mathbf{57.8_{\pm2.3}}$ & $\mathbf{58.3_{\pm2.8}}$ \\ \hline
100 & $82.5_{\pm3.5}$ & $56.3_{\pm1.8}$ & $57.2_{\pm2.4}$ \\ \hline
\end{tabular}
}

\vspace{.3em}

\resizebox{\textwidth}{!}{%
\begin{tabular}{|l|c|c|c|}
\hline
\multirow{2}{*}{ICs} & \multicolumn{3}{c|}{Mot} \\ \cline{2-4} 
 & $\text{MSE}_{adj}$ & MAE & $\text{MAE}_{adj}$ \\ \hline
25 & $95.7_{\pm9.2}$  & $0.76_{\pm0.04}$  & $7.7_{\pm0.3}$ \\ \hline
100 & $\mathbf{94.0_{\pm10.3}}$  & $\mathbf{0.75_{\pm0.03}}$ & $\mathbf{7.5_{\pm0.2}}$ \\ \hline
\hline
 & ACC & $\text{ACC}_{bal}$ & AUC \\ \hline
25 & $\mathbf{95.2_{\pm0.7}}$ & $55.3_{\pm3.6}$ & $56.1_{\pm3.8}$ \\ \hline
100 & $92.8_{\pm0.9}$ & $\mathbf{55.6_{\pm3.1}}$ & $\mathbf{56.7_{\pm3.4}}$ \\ \hline
\end{tabular}
}
\end{minipage}
\hspace{3em}
\begin{minipage}{0.4\textwidth}
\resizebox{\textwidth}{!}{%
\begin{tabular}{|l|c|c|c|}
\hline
\multirow{2}{*}{ICs} & \multicolumn{3}{c|}{Cog} \\ \cline{2-4} 
 & $\text{MSE}_{adj}$ & MAE & $\text{MAE}_{adj}$ \\ \hline
25 & $\mathbf{119.8_{\pm10.9}}$ & $\mathbf{0.61_{\pm0.06}}$ & $\mathbf{8.1_{\pm0.55}}$ \\\hline
100 & $121.1_{\pm7.0}$ & $0.65_{\pm0.07}$ & $8.6_{\pm0.8}$ \\ \hline
\hline
 & ACC & $\text{ACC}_{bal}$ & AUC \\ \hline
25 & $95.0_{\pm1.4}$ & $\mathbf{60.6_{\pm1.6}}$  & $\mathbf{62.2_{\pm1.6}}$ \\ \hline
100 & $\mathbf{95.8_{\pm1.3}}$ & $59.4_{\pm1.8}$ & $61.6_{\pm1.4}$ \\
\hline
\end{tabular}
}

\vspace{.3em}

\resizebox{\textwidth}{!}{%
\begin{tabular}{|l|c|c|c|}
\hline
\multirow{2}{*}{ICs} & \multicolumn{3}{c|}{Lang} \\ \cline{2-4} 
 & $\text{MSE}_{adj}$ & MAE & $\text{MAE}_{adj}$ \\ \hline
25 & $\mathbf{212.1_{\pm17.2}}$ & $\mathbf{0.85_{\pm0.02}}$ & $\mathbf{11.4_{\pm0.4}}$ \\ \hline
100 & $221.1_{\pm20.4}$ & $0.86_{\pm0.03}$  & $11.5_{\pm0.5}$ \\ \hline
\hline
 & ACC & $\text{ACC}_{bal}$ & AUC \\ \hline
25 & $\mathbf{83.9_{\pm2.6}}$  & $62.2_{\pm0.9}$  & $62.9_{\pm1.2}$ \\ \hline
100 & $81.3_{\pm3.3}$ & $\mathbf{62.7_{\pm1.3}}$  & $\mathbf{63.8_{\pm1.7}}$ \\ \hline
\end{tabular}
}

\vspace{.3em}

\resizebox{\textwidth}{!}{%
\begin{tabular}{|l|c|c|c|}
\hline
\multirow{2}{*}{ICs} & \multicolumn{3}{c|}{Mot} \\ \cline{2-4} 
 & $\text{MSE}_{adj}$ & MAE & $\text{MAE}_{adj}$ \\ \hline
25 & $\mathbf{90.1_{\pm9.6}}$ & $\mathbf{0.53_{\pm0.04}}$ & $\mathbf{7.0_{\pm0.5}}$ \\ \hline
100 & $90.3_{\pm10.5}$ & $0.55_{\pm0.04}$  & $7.1_{\pm0.6}$ \\ \hline
\hline
 & ACC & $\text{ACC}_{bal}$ & AUC \\ \hline
25 & $94.5_{\pm1.3}$  & $\mathbf{58.4_{\pm1.5}}$ & $\mathbf{60.1_{\pm1.7}}$ \\ \hline
100 & $\mathbf{95.0_{\pm1.5}}$ & $57.2_{\pm2.1}$ & $59.7_{\pm1.5}$ \\ \hline
\end{tabular}
}
\end{minipage}
\end{table}

\paragraph{Discussion}

This study demonstrates that integrating Group ICA-based dimensionality reduction with SwiFT significantly enhances predictions of neurodevelopmental outcomes from neonatal fMRI data. By extracting biologically meaningful features via ICA and leveraging multi-label learning, the approach preserves critical neural information while reducing computational complexity. SwiFT effectively learns local and global spatiotemporal patterns in 4D fMRI data, underscoring the synergy between neuroscience-driven feature engineering and advanced machine learning.
As can be seen in Table~\ref{tab:bsid-ica}, our results indicate that ICA dimensionality (25 vs. 100 ICs) does not substantially impact performance, suggesting that ICA robustly captures essential neural features regardless of the number of components. Furthermore, single-label learning with ICA-extracted features outperformed raw fMRI predictions, particularly for language classification and motor regression tasks, highlighting ICA’s ability to retain biologically meaningful information while reducing noise.
The combination of ICA and multi-label learning further enhanced predictive power, aligning with previous findings on the interrelated nature of developmental outcomes. These findings reinforce ICA’s value as a preprocessing step, enabling the model to focus on key neural networks and improve accuracy and efficiency. This targeted approach demonstrates the potential of integrating neuroscience-driven features with attention-based architectures for advancing neurodevelopmental research.

\subsection{Comparative Analysis}

\begin{figure}[h]
    \centering
    \begin{minipage}{0.45\textwidth}
        \centering
        \includegraphics[width=\textwidth]{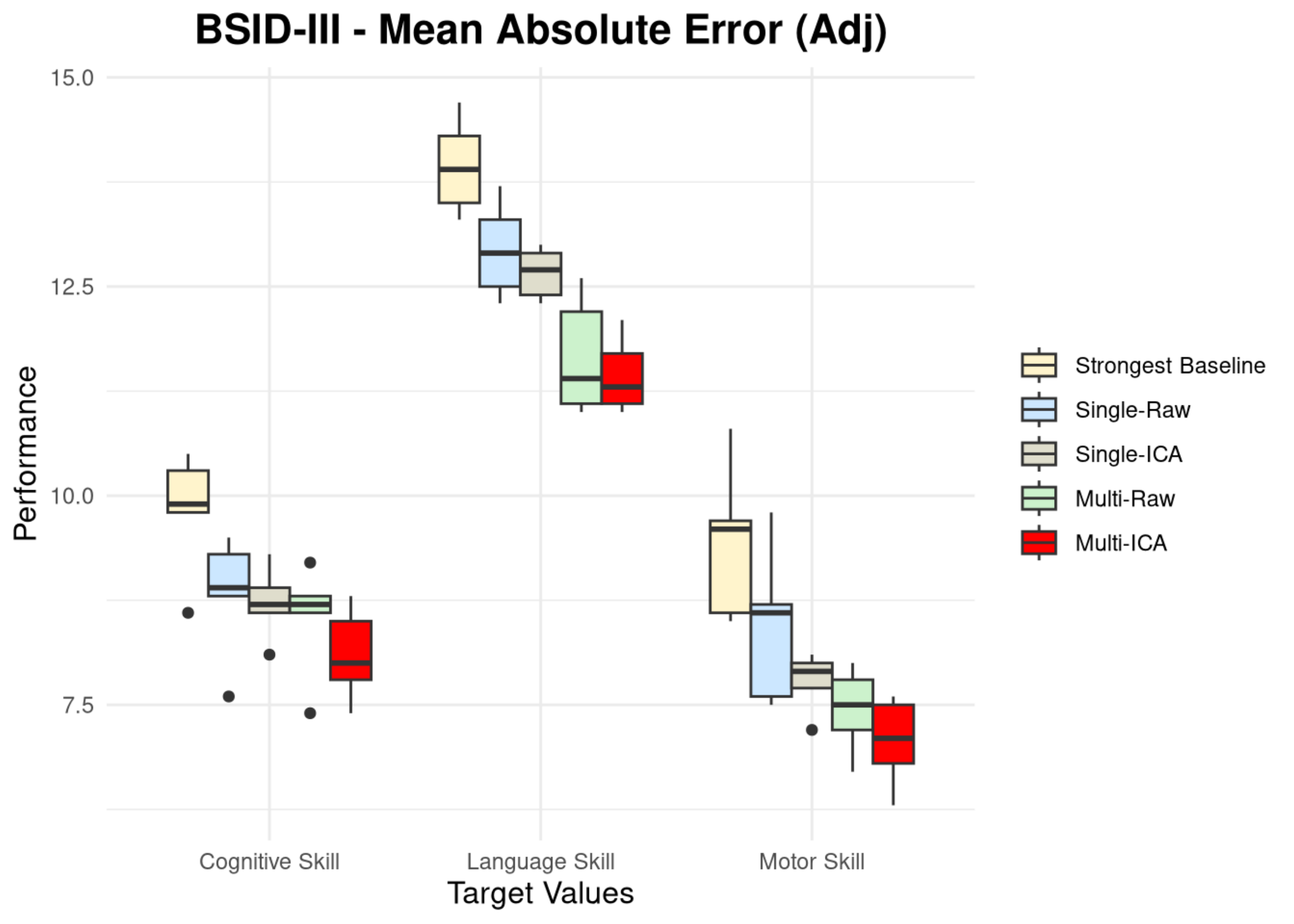}
    \end{minipage}
    \hfill
    \begin{minipage}{0.45\textwidth}
        \centering
        \includegraphics[width=\textwidth]{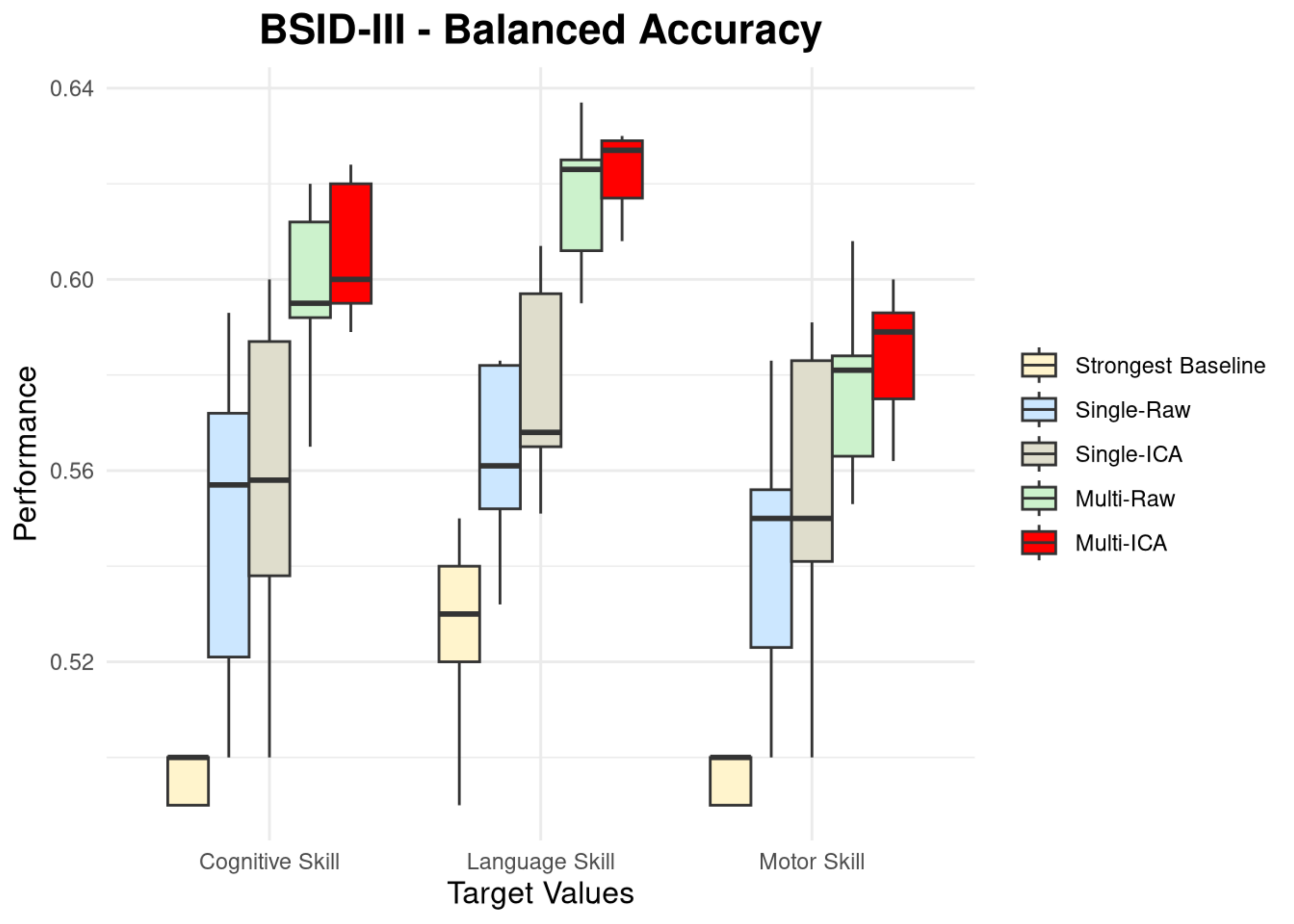}
    \end{minipage}
    \caption{Overview of the models' predictive power via 5-fold cross-validation for BSID-III regression ($\text{MAE}_{adj}$) and classification ($\text{ACC}_{bal}$). The strongest baseline was taken to represent baseline performance. Performance by transfer-learning and 100 ICs was omitted due to no statistically significant performance changes.}
    \label{fig:overview}
\end{figure}

We have summarized our experimental results via boxplots in Figure~\ref{fig:overview}, while omitting results obtained via transfer-learning and higher-dimensional ICs due to no statistically significant performance changes. After statistical significance analysis between the best-performing baseline model and SwiFT variation for each task across all metrics, Multi-ICA demonstrates significant and consistent improvements over the baseline model for most tasks, particularly in Cognitive MAE ($p = 0.004$) and Motor MAE ($p = 0.002$). 
Similarly, for regression tasks, significant reductions in cognitive regression $\text{MSE}_{adj}$ (\textit{p} = 0.036) and language regression $\text{MSE}_{adj}$ (\textit{p} = 0.008) highlight the robustness of the Multi-ICA approach in modeling complex developmental outcomes. Importantly, these improvements hold across multiple target domains, reinforcing the utility of a multi-label learning paradigm.
In classification, Multi-ICA also achieves significant improvements in language classification accuracy (\textit{p} = 0.004) and motor classification AUC (\textit{p} = 0.044), while maintaining competitive performance across most other metrics. However, some metrics -- such as 
cognitive classification AUC (\textit{p} = 0.413) -- did not show statistically significant differences compared to the baseline. These results suggest that the model’s performance in certain domains, particularly language-related tasks, may benefit from further domain-specific optimization.
Overall, Multi-ICA paired with multi-label learning consistently outperforms the strongest baseline model, confirming its utility for both regression and classification tasks in the context of predicting developmental outcomes.

\section{Interpretation via Integrated Gradients}
\label{sec:interpretations}

As in ~\cite{kim2023swiftswin4dfmri}, we used Integrated Gradient with Smoothgrad sQuare (IG-SQ), implemented through the Captum framework~\cite{captum1}. This method allowed us to identify spatial representations in brain regions that contribute to Bayley-III classification tasks. This analysis was based on the best-performing model, which combined multi-label prediction with ICA-extracted features. Leveraging ICs allowed for neuroscientifically meaningful interpretations by associating model outputs with known functional networks~\cite{BECKMANN2009S148}.
For each test sample, we generated 4D IG-SQ maps and excluded misclassified instances to ensure the validity of interpretations. The resultant maps were normalized, Gaussian-smoothed, and averaged across subjects to identify spatial patterns consistently associated with cognitive, language, and motor outcomes. Additionally, we aggregated these maps across ICs to derive comprehensive interpretations of brain activity relevant to each developmental domain.

\begin{figure}
    \centering
    \includegraphics[width=0.6\linewidth]{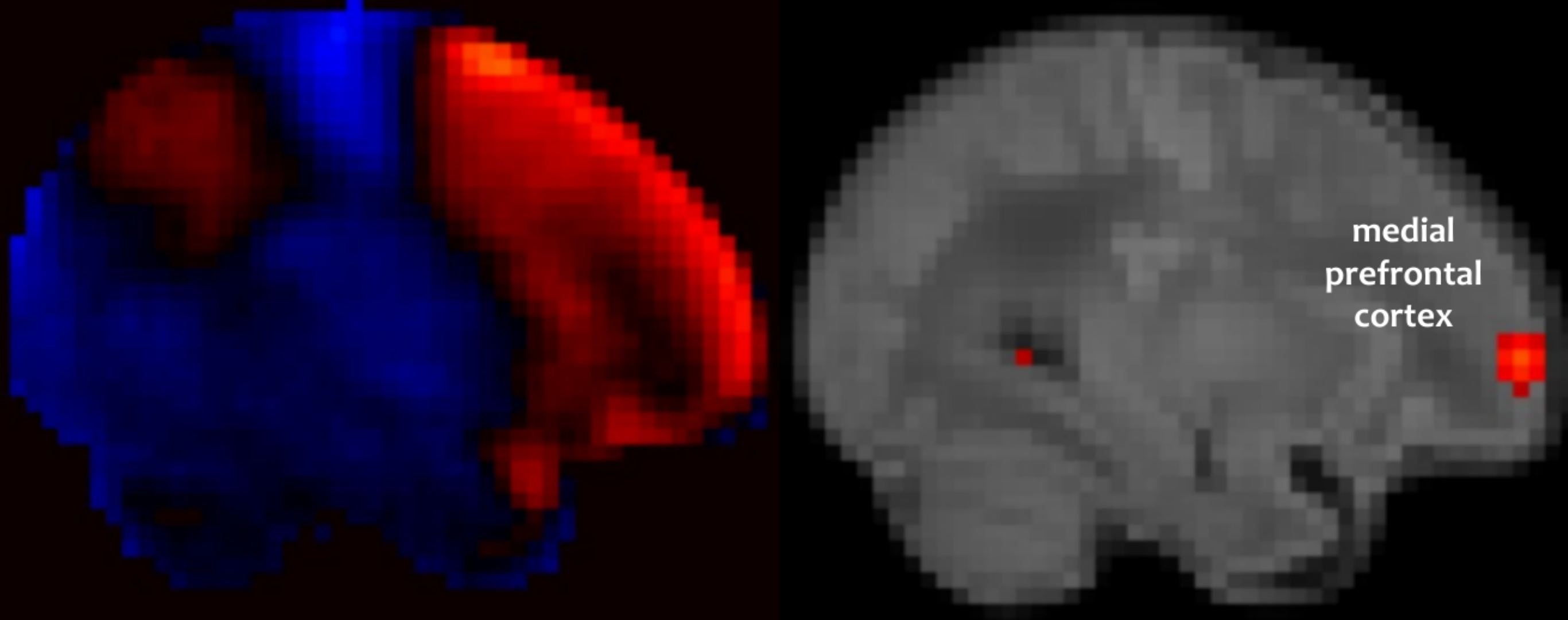}
    \caption{A glimpse into network-level interpretations, focusing on IC~5~(left) possibly representing the Executive Control Network~(ECN). We can see how this network adds to activation in the medial prefrontal cortex for prediction of cognitive delay (right).}
    \label{fig:ica_int}
\end{figure}

\begin{figure}[h]
    \centering
    \begin{minipage}{0.6\linewidth}
        \centering
        \includegraphics[width=\textwidth]{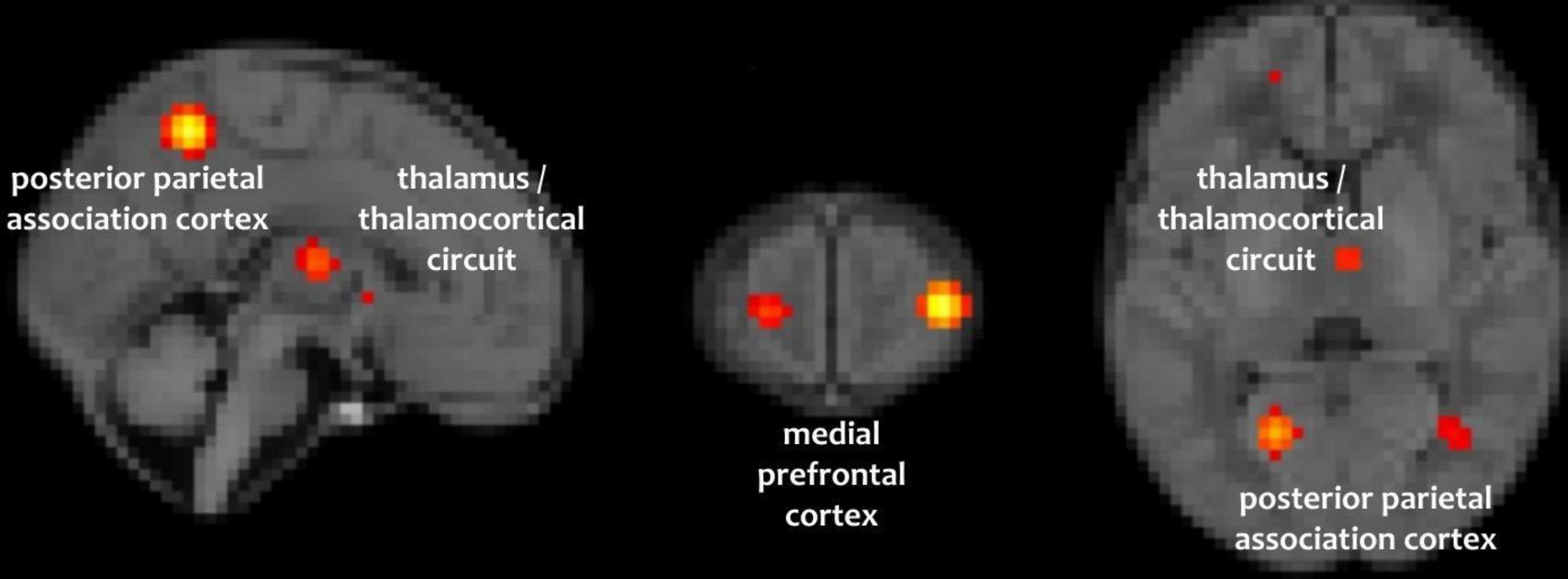}
    \end{minipage}
    
    \vspace{0.2em}
    
    \begin{minipage}{0.6\linewidth}
        \centering
        \includegraphics[width=\textwidth]{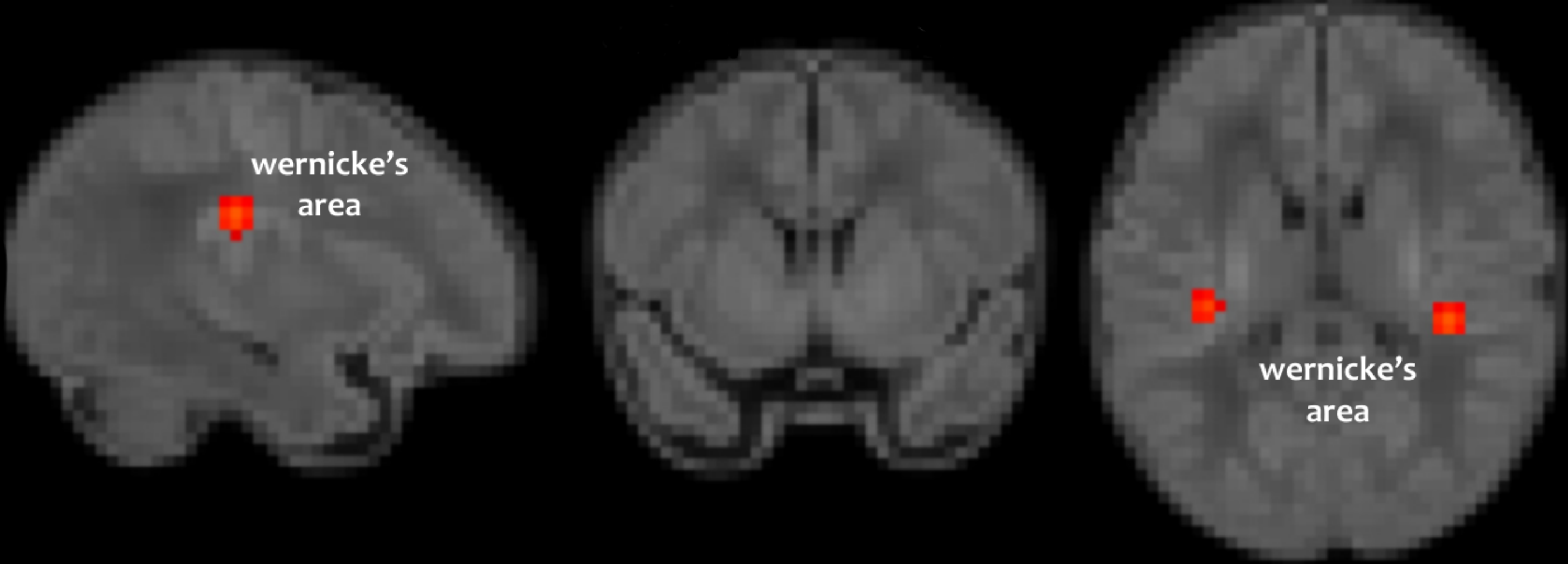}
    \end{minipage}
    
    \vspace{0.2em}
    
    \begin{minipage}{0.6\linewidth}
        \centering
        \includegraphics[width=\textwidth]{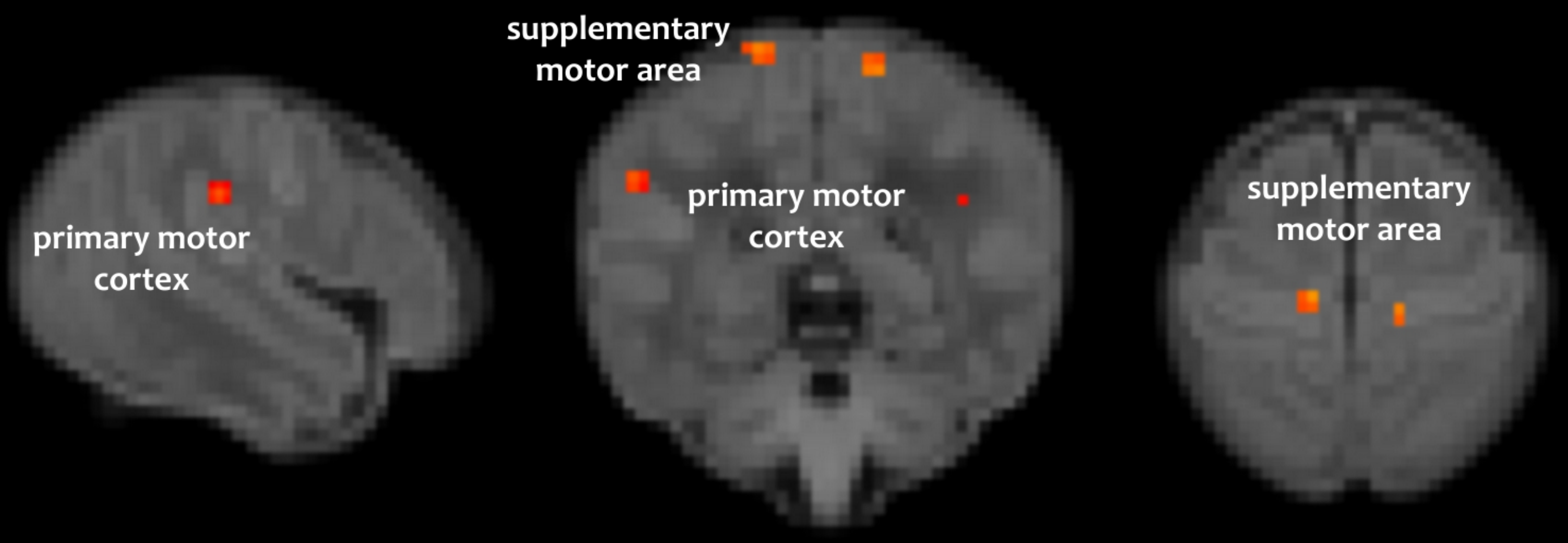}
    \end{minipage}
    \caption{Interpretation maps with thresholded Integrated Gradients (IG) maps averaged across ICs for risk of cognitive (top), language (middle) and motor (bottom) delay.}
    \label{fig:interpretation}
\end{figure}

The analysis revealed distinct brain regions whose spatiotemporal dynamics contribute to each developmental outcome (Figure~\ref{fig:interpretation}). For cognitive delay risk, activations were observed in the medial prefrontal cortex (coronal view), thalamocortical circuit (sagittal and axial view), and posterior parietal association cortex (sagittal view). These findings align with neuroscience literature: the identified brain regions are key to early sensory processing, attention, executive function, and spatial cognition, all foundational to normal cognitive development~\cite{mpfc}~\cite{thalamus}~\cite{ppc}. 
In Figure~\ref{fig:ica_int}, network-level predictions are further explored. For instance, IC5 likely represents the Executive Control Network (ECN)\cite{Li2019}, a critical network for cognitive functions such as attention and decision-making. Its contribution to the overall interpretation, as shown in Figure~\ref{fig:interpretation}, is evident through activations in the medial prefrontal cortex (mPFC), a hallmark region of the ECN.
For the risk of linguistic delay (middle), activations in Wernicke's area align with its well-documented role in language comprehension and development~\cite{Binder2015}. Similarly, for motor skill delays (bottom), activations in the primary motor cortex and supplementary motor areas (SMAs) highlight their roles in gross and fine motor control and motor pattern learning, respectively~\cite{Graziano2006}~\cite{Tanji1994}. 

These findings reinforce the model's ability to capture neurobiologically meaningful spatiotemporal patterns, as discussed earlier. By integrating both spatial localization and temporal dynamics of brain activity, the model effectively associates developmental delays with functional neural circuits. This alignment with established neuroscience underscores the robustness of our approach, demonstrating how machine learning can provide interpretable insights into the neural basis of developmental outcomes.

\section{Concluding Remarks}
\label{sec:conclusion}

In this study, we demonstrated that SwiFT provides a significant improvement in evaluating neonatal fMRI data for predicting early neurodevelopmental outcomes. By integrating multi-label learning and leveraging ICA-extracted features, we achieved enhanced predictive accuracy while improving model interpretability. These advances suggest that SwiFT has the potential to play a pivotal role in the early detection of developmental delays, paving the way for personalized therapeutic interventions for at-risk neonates. The clinical relevance of such a model is strengthened by a study suggesting that therapeutic interventions to treat neurodevelopmental disorders may be more effective if done during the early stages of brain development~\cite{SVALINA2022}.
Despite these accomplishments, limitations remain. The imbalanced nature of the dataset poses a  challenge, impacting the reliability of the classification tasks. While our approach partially mitigated this issue, future work should explore advanced strategies for handling data imbalance, such as oversampling or even synthetic data generation. Additionally, further validation of ICA-extracted features as proxies for brain-network-level mechanisms is necessary to strengthen the biological interpretability of our findings. Also, expanding this framework to include datasets from other age groups, such as toddlers, children, and adults, could enhance the generalizability of the model. Extending the multi-label learning paradigm to incorporate additional target variables, such as the Q-CHAT score for early autism screening, offers an exciting direction for future research and could be implemented into the current pipeline without major changes Finally, pretraining on adult data could provide a robust foundation for the model, but has shown no generalizability to neonates in our experiments. Since the use of small neonatal datasets increases the risk of overfitting, examining other pretraining paradigms than contrastive learning, such as Masked Image Modeling, may be beneficial.

In conclusion, this work establishes a robust foundation for advancing predictive and interpretable models of neurodevelopment. With continued refinement and access to diverse, large-scale datasets, SwiFT holds significant potential for innovations in neuroscience and personalized medicine.


\section*{Acknowledgements}
\label{sec:acknowledgements}
{
This work was supported by the National Research Foundation of Korea(NRF) grant funded by the Korea government(MSIT) (No. 2021R1C1C1006503, RS-2023-00266787, RS-2023-00265406, RS-202400421268), by Creative-Pioneering Researchers Program through Seoul National University(No. 20020240057), by Semi-Supervised Learning Research Grant by SAMSUNG(No.A0426-20220118), by Identify the network of brain preparation steps for concentration Research Grant by LooxidLabs(No.33920230001), by Institute of Information \& communications Technology Planning \& Evaluation (IITP) grant funded by the Korea government(MSIT) [NO.RS-2021-II211343, Artificial Intelligence Graduate School Program (Seoul National University)] by the MSIT(Ministry of Science, ICT), Korea, under the Global Research Support Program in the Digital Field program(RS-2024-00421268) supervised by the IITP(Institute for Information \& Communications Technology Planning \& Evaluation), by the National Supercomputing Center with supercomputing resources including technical support(KSC-2023CRE-0568) and by the Ministry of Education of the Republic of Korea and the National Research Foundation of Korea (NRF - 2021S1A3A2A02090597), and by Artificial intelligence industrial convergence cluster development project funded by the Ministry of Science and ICT(MSIT, Korea) \& Gwangju Metropolitan City. This research used resources of: the Oak Ridge Leadership Computing Facility, a DOE Office of Science User Facility supported under Contract DE-AC05-00OR22725; the National Energy Research Scientific Computing Center, a DOE Office of Science User Facility using NERSC award NERSC DDR-ERCAP0030592 and DDR-GenAI, and ALCC-ERCAP0030659.
}

\end{document}